# Cost-Effectiveness of Adult Hepatitis A Vaccination Strategies in Korea Under an Aging Susceptibility Profile

Yuna Lim[1], Gerardo Chowell[2], and Eunok Jung[1]*

[1] Department of Mathematics, Konkuk University, Seoul, Republic of Korea

[2] Department of Population Health Sciences, Georgia State University, Atlanta, United States

This is information for the corresponding author:

Name: **Eunok Jung**

    E-mail: junge@konkuk.ac.kr

    ORCID: 0000-0002-7411-3134

    Address: Department of Mathematics, Konkuk University

        120, Neungdong-ro, Gwangjin-gu, Seoul, 143-701, Republic of Korea

    Tel: +82-10-9973-4163

    Fax: +82-2-458-1952

**Abstract**

Background: Hepatitis A severity increases sharply with age, while Korea is experiencing a cohort shift in which low-seroprevalence adult cohorts are aging into older, higher-fatality age groups. This demographic and immunological transition creates an urgent policy question regarding how adult vaccination should be prioritized under resource constraints.

Methods: We evaluated three adult vaccination scenarios targeting low-seroprevalence age groups: ($S1$) 20–39 years, ($S2$) 40–59 years with pre-vaccination antibody testing, and ($S3$) 20–59 years with antibody testing applied to those aged ≥40 years. Using an age-structured dynamic transmission model calibrated to Korean data, we derived dynamically feasible vaccination allocation trajectories under realistic capacity constraints using an optimal control framework and linked these trajectories to long-term transmission-model simulations. We conducted DALY-based cost-effectiveness analyses over a lifetime horizon from both healthcare system and societal perspectives, and characterized uncertainty using probabilistic sensitivity analysis (PSA) and cost-effectiveness acceptability curves (CEACs). Robustness was examined using one-way sensitivity analyses.

Results: In the base case (2025 USD; 4.5% annual discount rate), vaccination targeting adults aged 40–59 years ($S2$) consistently yields the most favorable and robust cost-effectiveness profile under both perspectives, with the lowest ICER. Vaccination of adults aged 20–59 years ($S3$) achieved the largest reduction in DALYs but requires substantially higher incremental costs, resulting in a higher ICER than $S2$. $S1$ produces the smallest DALY reduction and is the least efficient strategy. PSA and CEACs confirm that $S2$ remains the preferred option across most willingness-to-pay ranges.

Conclusions: Prioritizing adults aged 40–59 years offers the most balanced and robustly cost-effective strategy in Korea, capturing substantial mortality reduction while limiting additional program costs. Broader adult vaccination may be justified when higher budgets or willingness-to-pay thresholds are acceptable, but prioritizing middle-aged adults provides the clearest value for money under epidemiological and economic conditions.

**Key points for decision makers**

- In Korea, hepatitis A susceptibility is shifting toward older adults, among whom disease severity and mortality risk increase sharply, underscoring the need for targeted adult vaccination policies.
- Vaccinating adults aged 40–59 years with pre-vaccination antibody testing is the most robustly cost-effective option under realistic program capacity constraints.
- Vaccinating a broader adult age range (20–59 years) can avert a larger overall disease burden, but requires substantially greater investment and is most likely to be preferred when decision-makers accept higher budgets or willingness-to-pay levels.

**Introduction**

Hepatitis A is caused by infection with the hepatitis A virus (HAV) and is transmitted not only via contaminated food or water but also through direct person-to-person contact. Although often self-limiting in younger individuals, disease severity increases with age. The case fatality rate is reported to be approximately 0.2% overall but rises to about 1.8% among adults aged ≥50 years [1,2]. This strong age gradient in severity implies that population aging and shifts in immunity profiles can substantially alter the public health impact of hepatitis A. Hence, hepatitis A poses a disproportionate mortality risk in older adults, raising an important and timely policy question regarding how limited preventive resources should be allocated across adult age groups.

In Korea, hepatitis A seroprevalence exhibits marked heterogeneity by age. As of 2019, seroprevalence was lowest among individuals in their teens through their 40s, and a cohort shift is underway, whereby low-immunity cohorts are progressively aging into older groups [3]. In contrast, relatively high seroprevalence among adults aged ≥50 years is interpreted as immunity acquired through natural infection during childhood or adolescence when sanitary conditions were poorer. As hygiene improved, individuals born after the 1970s experienced reduced HAV exposure in childhood and therefore did not acquire sufficient natural immunity. Since the introduction of the hepatitis A vaccine in 1997 and the incorporation of hepatitis A vaccination into the National Immunization Program in 2015 (free vaccination for children born on or after January 1, 2012), seroprevalence in adolescents has increased, while susceptibility has accumulated among unvaccinated adults [3]. As a result, Korea is undergoing a demographic and immuno-epidemiological transition in which susceptibility is shifting toward older age groups with higher fatality risk.

Recent age-specific case distributions suggest that hepatitis A infection is no longer confined to adults aged 20–49 years but has expanded to those aged ≥50 years, indicating that low-immunity patterns may also be emerging in older age groups. This cohort-driven shift implies that future hepatitis A burden will increasingly concentrate in age groups with substantially higher fatality risk, sharpening the policy trade-off between targeting younger adults with higher incidence versus older adults with higher mortality. Combined with the sharp increase in case fatality at older ages, this ongoing immuno-epidemiological transition could substantially increase future mortality attributable to hepatitis A in older adults. Moreover, because adults aged 20–59 years constitute the core working-age population, infections in these groups can lead to substantial productivity losses due to treatment and isolation. Accordingly, hepatitis A burden should be evaluated not only in terms of incidence and mortality, but also in terms of broader social and economic consequences. This raises the key question of whether adult vaccination strategies should prioritize younger working-age adults to reduce productivity losses or middle-aged adults to avert severe outcomes and deaths.

Because hepatitis A is effectively preventable through vaccination and additional adult vaccination is currently not implemented in Korea except for selected groups such as military personnel, the central policy question is not whether to vaccinate adults, but which adult age groups should be prioritized under realistic resource and implementation constraints. However, rigorous evidence in Korea remains limited on the cost-effectiveness of adult vaccination strategies that reflect the current epidemiological context while jointly considering long-term health outcomes and associated socioeconomic costs. In particular, there is a lack of economic evaluations that explicitly incorporate cohort-driven aging of susceptibility and age-dependent disease severity within a dynamic transmission framework.

Globally, economic evaluations of hepatitis A vaccination have largely focused on routine childhood immunization, catch-up programs, or selected high-risk groups, and evidence for adult age-targeted strategies in high-income settings remains relatively limited and context dependent. For example, Dhankhar et al. used an age-structured dynamic transmission model for the United States and reported that universal childhood vaccination was cost-saving compared with a regional policy once herd protection was incorporated [4]. Elbasha et al. similarly applied dynamic transmission modeling to evaluate routine vaccination plus catch-up vaccination (ages 2–18 years) and found catch-up to be cost-saving over a long horizon by closing immunity gaps [5]. In Korea, a national research report commissioned by the Korea Disease Control and Prevention Agency (KDCA; formerly the Korea Centers for Disease Control and Prevention) in 2009 conducted a long-horizon economic evaluation of hepatitis A vaccination options using QALYs [6]. More recently, Lim et al. (arXiv preprint) developed an age-structured transmission model to compare adult vaccination strategies targeting the 20s–30s versus the 40s–50s and evaluated strategies primarily in terms of projected deaths under fixed cost or supply constraints [7]. However, prior studies have generally not combined cohort-shifted susceptibility, transmission dynamics, and formal long-term cost-effectiveness analysis across adult age groups within a unified framework.

In this study, we address this gap by developing an age-structured transmission model calibrated to Korean age-specific epidemiological data and conducting a long-term, simulation-based cost-effectiveness analysis of adult hepatitis A vaccination strategies from both the healthcare system and societal perspectives. The analysis explicitly incorporates cohort-driven aging of susceptibility, sharply age-dependent fatality risk, productivity losses, and multiple sources of uncertainty. By jointly integrating transmission dynamics and economic evaluation, this work provides policy-relevant evidence to inform prioritization of adult vaccination under realistic capacity and budget constraints in Korea.

## Materials and Methods

**Mathematical model and scenarios**

The population is stratified into eight 10-year age groups: $G_0$(0–9), $G_1$(10–19), $G_2$(20–29), $G_3$(30–39), $G_4$(40–49), $G_5$(50–59), $G_6$(60–69), and $G_7$(≥70 years). An age-structured hepatitis A transmission model incorporating vaccination controls is shown in Figure 1, and the full system of equations and parameter definitions are provided in Supplementary Appendix 1. The model is designed to capture cohort-driven changes in susceptibility, age-dependent disease severity, and realistic vaccination implementation features relevant to the Korean context.

Susceptible individuals ($S$) transition to the exposed state ($E$) at a rate determined by the force of infection, $\lambda(t)$. Following an average latent period of $1/\kappa$ days, exposed individuals progress to the infectious stage and are partitioned into symptomatic infectious ($I$) and asymptomatic infectious ($A$) individuals, with proportions of $p$ and $1-p$, respectively. Symptomatic infectious individuals ($I$) develop jaundice-like symptoms after an average of $1/\alpha$ days and are subsequently transitioned to the isolated group ($Q$). Following an average isolation period of $1/\gamma$ days, isolated individuals either recover with awareness of infection ($R_a$) or die ($D$), with age-dependent (two-tier) case-fatality rate (CFR) $f$. Asymptomatic infectious individuals ($A$) recover without awareness of prior infection ($R_u$) after an average infectious period of $1/\eta$ days. This structure allows the model to distinguish between clinically apparent and inapparent infections and to represent the sharp increase in fatality risk at older ages.

In this study, our model incorporates Korea's National Immunization Program (NIP) to reflect real-world vaccination dynamics. In $G_0$, monthly births ($\Lambda$) enter either $S_0$ or the first-dose vaccinated individuals $V_0$ according to the infant vaccination coverage rate, $\nu$. First-dose vaccinated individuals ($V$) receive a second dose after an average interval between the first and the second dose of $1/\rho$, and transition to $R_a$ or they may still be infected with residual susceptibility determined by the vaccine effectiveness of first dose, $e$. To reflect vaccination for military recruits, we additionally include first-dose vaccinated individuals $V_M$ in $G_2$. Individuals in $V_M$ who complete the second dose move to a protected class (denoted $R_{a,2}$). Those who do not receive the second dose may lose protection after duration of a single dose effectiveness and return to susceptibility in the subsequent age group $S_3$, or may become infected while partially protected. With monthly time steps and 10-year age groups, the ageing rate is set to $1/(12 \times 10)$ per month. Subscripts $i$ and $j$ indicate age groups in model flowchart. The vaccination control variables shown in red in Figure 1 represent the monthly number of individuals vaccinated in each adult age group.

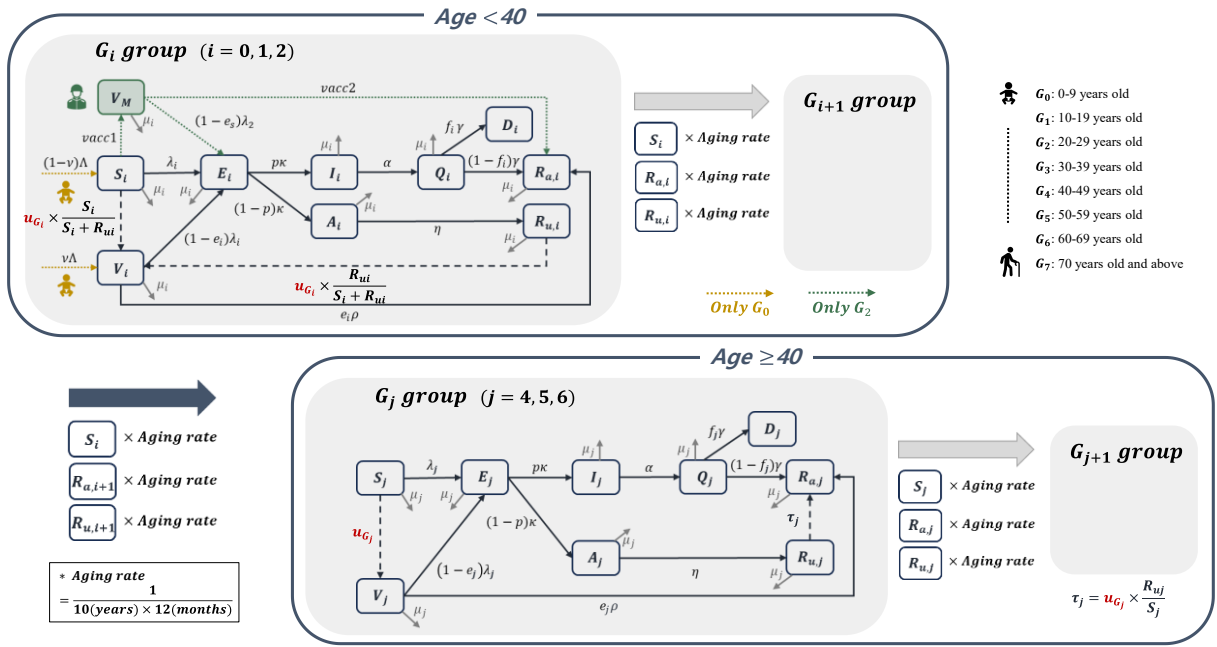

**Fig. 1** Age-structured hepatitis A transmission model flowchart with vaccination controls. Compartments represent epidemiological states stratified by age, and arrows indicate transitions between states. Full model equations and parameter definitions are provided in Supplementary Appendix 1

Given that Korea mandates pre-vaccination antibody testing for adults aged ≥40 years, we incorporate two distinct adult vaccination structures into the model. For age groups <40 years (indexed by $i$), vaccination is administered without prior testing. Consequently, both susceptible individuals ($S_i$) and recovered individuals without awareness of infection ($R_{u,i}$) are eligible for vaccination. We allocate the administered doses proportionally to the sizes of $S_i$ and $R_{u,i}$. For age groups ≥40 years (indexed by $j$), antibody testing is conducted before vaccination. As a result, vaccination is administered only to susceptible individuals ($S_j$), while individuals in $R_{u,j}$ who are screened out by testing are reclassified as recovered individuals with awareness of infection ($R_{a,j}$) (Supplementary Appendix 1). This structure reflects Korea's current vaccination practice and ensures that programmatic costs associated with antibody testing are appropriately represented.

Building on this model, we analyze three adult vaccination scenarios targeting age groups with low seroprevalence:

*Scenario* **1** (*S***1**) vaccination of adults aged 20–39 years without prior testing,

*Scenario* **2** (*S***2**) vaccination of adults aged 40–59 years with pre-vaccination antibody testing,

*Scenario* **3** (*S***3**) vaccination of adults aged 20–59 years with pre-vaccination antibody testing applied to those aged ≥40 years.

The baseline scenario is defined as no additional adult vaccination beyond existing practice. These scenarios were selected to reflect realistic policy options currently under consideration in Korea and to isolate the trade-offs between incidence reduction, mortality prevention, and vaccination program costs. All scenarios are evaluated under identical transmission and economic assumptions, differing only in the targeted age groups and testing requirements.

**Data**

We use age-specific monthly hepatitis A confirmed case data provided by the KDCA. The surveillance data are reported in 10-year age groups (0–9, 10–19, 20–29, 30–39, 40–49, 50–59, 60–69, and ≥70 years) and cover the period from January 2016 through December 2024. To initialize age-specific immunity, we use age-specific anti-HAV seroprevalence values from the 2015 National Health Statistics published by the Ministry of Health and Welfare, reported in the same 10-year age groups [8]. These seroprevalence data provide cohort-specific initial conditions reflecting historical differences in exposure and vaccination uptake across birth cohorts. Together, the age-specific incidence and seroprevalence data allow us to jointly initialize cohort-driven susceptibility patterns and calibrate the force of infection across age groups, ensuring consistency between observed epidemiological trends and modeled transmission dynamics.

**Parameter estimation**

The age-specific force of infection $\lambda_k(t)$ combines person-to-person transmission and an external component representing foodborne/waterborne infection, with additional event term capturing the contaminated salted-clam outbreaks in 2019 and 2021 (Supplementary Appendix 3). This formulation allows the model to reproduce both sustained endemic transmission and sharp, short-term outbreak-driven increases in incidence observed in the data. Key parameters in $\lambda_k(t)$, including event-related intervention effects ($\xi_1$ and $\xi_2$), are estimated via the least squares method using MATLAB (lsqcurvefit function). Prior to parameter estimation, we assess structural identifiability of model parameters using a reduced single-population (age-aggregated) version of the model implemented in Julia (Supplementary Appendix 2) [9,10]. Parameter uncertainty is quantified via bootstrap resampling (100,000 replicates) [11], from which we report 95% bootstrap confidence intervals and generate 95% predictive intervals for age-specific cumulative cases (Supplementary Appendix 3).

Although the reduced model is globally structurally identifiable under the specified output–input setting (Supplementary Table 2), structural identifiability alone does not guarantee narrow confidence intervals when the model is confronted with real limited data. To account for this uncertainty, all downstream health and economic outcomes are evaluated using probabilistic sensitivity analysis, ensuring that policy conclusions reflect both structural and practical sources of uncertainty.

**Optimal control framework for resource allocation**

For each scenario, we employ optimal control theory to derive monthly vaccination allocation trajectories for the period 2025-2054, incorporating realistic implementation constraints. The objective of this framework is to identify dynamically feasible and efficient prioritization patterns across adult age groups under limited vaccination capacity, rather than to prescribe exact operational schedules. The objective functional, $J$, is defined to minimize the total cost (disease burden and socioeconomic costs of the vaccination program) over the planning horizon as follows:

$$J(u_{G_2}, u_{G_3}, u_{G_4}, u_{G_5})$$
$$= \int_0^{360} C_{Death} \sum_{k=0}^{7} \gamma f_k Q_k(t) + \sum_{k=0}^{7} C_{Illness}^k p\kappa E_k(t) + C_{Vacc} \sum_{l=2}^{5} u_{G_l}(t) + C_{Test} \sum_{m=4}^{5} T_m(t)$$
$$+ \frac{w}{2} \sum_{l=2}^{5} u_{G_l}^2(t) \, dt.$$

Here, $u_{G_l}(t)$ denotes the monthly number of first-dose vaccinations administered in adult age group $l \in \{2,3,4,5\}$. For age groups $m = 4,5$ (40-59 years), vaccination is preceded by antibody testing, so the total number of tests is $T_m(t) = u_{G_m}(t) + \tau_m(t)$. Controls are subject to a monthly capacity constraint and a scenario-specific isoperimetric constraint on total vaccinations (Supplementary Appendix 4) [12]. In particular, the monthly capacity constraint is modeled to represent the feasible uptake under routine (non-outbreak) conditions, reflecting expected demand and adherence rather than vaccine supply limitations. A quadratic penalty term for the vaccination controls, $\frac{w}{2}\sum_{l=2}^{5} u_{G_l}^2(t)$, is incorporated to prevent extreme allocations and improve numerical stability [13,14]. The weight $C_{Illness}^k$ represents age-specific socioeconomic burden per symptomatic case, including medical costs and productivity losses associated with illness. All cost weights are normalized using the unit antibody testing cost (Supplementary Table 4). Using the scenario-specific optimal vaccination trajectories, we simulate the transmission model starting from 2025 over the evaluation horizon (83 years, based on average life expectancy in Korea) to compute long-term DALYs for each scenario and for the baseline [15]. Additional adult vaccination is set to zero after the 30-year control period (2025−2054). Time-series projections of symptomatic cases, deaths, and age-specific seroprevalence over the evaluation horizon are presented in Supplementary Figure 2.

The purpose of the optimal control framework is not to prescribe exact month-by-month vaccination schedules for implementation, but rather to identify dynamically feasible and efficient allocation patterns under realistic capacity and budget constraints. The resulting vaccination trajectories are therefore interpreted as policy-relevant guidance on prioritization across age groups over time.

**Calculation of DALYs**

Disability-adjusted life years (DALYs) are defined as the sum of years of life lost (YLL) due to premature mortality and years lived with disability (YLD). We compute age-specific contributions to YLD and YLL over the evaluation horizon using the monthly flows of symptomatic infections and deaths generated by the transmission model. Total DALYs are obtained by summing these values across all age groups. Time ($t$) is expressed in months from the start of the evaluation horizon ($t \in [t_0, t_f]$).

Using a continuous annual discount rate $r$, we apply the discount factor $\exp(-rt/12)$. The discounted average duration of illness is $AD = (1 - e^{-rID})/r$, where $ID$ is the mean illness duration (years). The discounted remaining life expectancy for age group $k$ is $L_k = (1 - e^{-rLE_k})/r$, where $LE_k$ is the remaining life expectancy (years). Remaining life expectancy by age group is obtained from the 2023 Korean life table [15].

The discounted YLD and YLL for all age groups are computed as follows:

$$\text{YLD} = \sum_k \left( \int_{t_0}^{t_f} p\, \kappa E_k(t)\, e^{-rt/12}\, dt \right) \times DW \times AD,$$

$$\text{YLL} = \sum_k \left( \int_{t_0}^{t_f} \gamma\, f_k Q_k(t)\, e^{-rt/12}\, dt \right) \times L_k,$$

where DW denotes the disability weight for hepatitis A obtained from the Global Burden of Disease Study 2013 estimates [16]. Note that for each age group $k$, the model generates symptomatic infections $p\kappa E_k(t)$ and deaths $\gamma f_k Q_k(t)$.

It is worth noting that the choice of a lifetime evaluation horizon and an annual discount rate of 4.5% follows the most recent Korean pharmacoeconomic evaluation guidelines and reflects standard practice for health technology assessment in Korea. These assumptions are therefore intended to align the analysis with national decision-making contexts rather than to represent generic international defaults.

**Estimation of age-specific costs: healthcare system and societal perspectives**

The economic evaluation is conducted from both the healthcare system and the societal perspectives, with costs categorized into vaccination program and treatment related expenditures. From the healthcare system perspective, vaccination program costs include direct costs such as vaccine prices, administration fees, and antibody testing costs, while treatment costs are limited to direct medical costs. From the societal perspective, vaccination program incorporates additional indirect costs, such as productivity losses associated with time spent visiting healthcare facilities and transportation costs. Similarly, treatment costs are expanded to include both direct medical and non-medical costs, along with all relevant indirect costs, including productivity losses due to illness and premature mortality. This dual-perspective approach allows us to distinguish between budgetary impacts on the healthcare system and broader economic consequences borne by society.

When source years differ, costs are converted to 2025 values using the consumer price index (CPI) and converted to US dollars using the August 2025 KRW–USD exchange rate (1,385 KRW per USD). Costs and health outcomes are discounted at 4.5% annually, following the updated Korean pharmacoeconomic evaluation guideline (third version) [17]. Applying a consistent discount rate across costs and outcomes ensures comparability with health technology assessments conducted in Korea. Cost inputs are derived from nationally representative Korean data sources (e.g., Statistics Korea, Ministry of Employment and Labor, KNHANES, HIRA, and NHIS) [6,18–23]. The complete costing equations and assumptions are provided in Supplementary Appendix 5. Perspective-specific cost components and aggregation are summarized in Tables 1–2. All costs are calculated at the age-group level, ensuring consistency with the age-structured transmission model and allowing age-specific differences in clinical pathways, healthcare utilization, and productivity losses to be reflected.

**Table 1** Age-specific costs for healthcare system perspective (Unit: US dollars). Costs include direct vaccination program costs (vaccine acquisition, administration, and antibody testing where applicable) and direct medical treatment costs. All values are discounted at an annual rate of 4.5%.

| | Cost | | | |
|---|---|---|---|---|
| | **Vaccination cost** | | **Treatment cost** | **Total cost** |
| | **Direct cost** | | **Direct cost** | |
| **Age** | Vaccine cost + Administration cost | Antibody test cost | Medical cost | |

| Age | | | | |
|---|---|---|---|---|
| 0-9 | - | - | 1106.54 | 1106.54 |
| 10-19 | - | - | 1613.52 | 1613.52 |
| 20-29 | 75.65 | - | 2226.89 | 2302.54 |
| 30-39 | 75.65 | - | 2136.45 | 2212.10 |
| 40-49 | 75.65 | 14.26 | 5106.08 | 5195.99 |
| 50-59 | 75.65 | 14.26 | 917.37 | 1007.28 |
| 60-69 | - | - | 860.63 | 860.63 |
| 70+ | - | - | 1646.37 | 1646.37 |

**Table 2** Age-specific cost for societal perspective (Unit: US dollars). Costs include direct medical and non-medical costs as well as indirect productivity losses due to illness, healthcare visits, and premature mortality. All values are discounted at an annual rate of 4.5%.

| | Cost | | | | | | Total cost |
|---|---|---|---|---|---|---|---|
| | **Vaccination cost** | | **Treatment cost** | | | | |
| | **Direct cost** | **Indirect cost** | **Direct cost** | | **Indirect cost** | | |
| Age | Vaccine cost + Administration cost + transportation costs | Antibody test cost | Productivity loss due to time spent visiting healthcare facilities | Medical cost | Non-medical cost (Transport + Supplemental care) | Productivity loss due to treatment | Productivity loss due to death | |
| 0-9 | - | - | 5.72 | 1234.98 | 21.47 | 0.00 | 316617.13 | 317879.30 |
| 10-19 | - | - | 3.19 | 1799.36 | 147.89 | 15.09 | 541944.92 | 543910.45 |
| 20-29 | 77.68 | - | 8.31 | 2482.18 | 13.49 | 1068.13 | 658307.97 | 661957.76 |
| 30-39 | 77.68 | - | 16.07 | 2381.50 | 105.89 | 2001.85 | 652364.82 | 656947.81 |
| 40-49 | 77.68 | 14.26 | 18.88 | 5690.95 | 271.76 | 1876.97 | 524015.18 | 531965.68 |
| 50-59 | 77.68 | 14.26 | 17.47 | 1024.37 | 530.57 | 958.18 | 292777.63 | 295400.16 |
| 60-69 | - | - | 9.76 | 961.19 | 564.26 | 237.93 | 26652.75 | 28425.89 |
| 70+ | - | - | 5.72 | 1835.85 | 523.78 | 0.00 | 0 | 2365.35 |

**Cost-effectiveness analysis**

We compare each adult vaccination scenario against the baseline. The incremental cost effectiveness ratio (ICER) is defined as

$$\text{ICER}_{i \text{ vs base}} = \frac{\text{Cost}_i - \text{Cost}_{base}}{\text{DALY}_{base} - \text{DALY}_i},$$

so that the denominator represents DALYs averted by scenario $i$. We first compute total costs, DALYs, and ICERs in the base-case analysis, in which all uncertain inputs are fixed at their reference values, for each scenario relative to the baseline. We present probabilistic sensitivity analysis (PSA) results on the cost-effectiveness plane to visualize joint uncertainty in incremental costs and DALYs averted. Reference ICER lines (e.g., 1× and 3× GDP per capita; USD/DALY) are displayed for illustrative purposes only, as Korea has no official willingness-to-pay (WTP) threshold [24,25]. Cost-effectiveness acceptability curves (CEACs) are constructed from PSA to show, across a broad range of WTP values, the probability that each scenario is the most cost-effective option under parameter uncertainty.

**Uncertainty and sensitivity analyses**

We perform both deterministic sensitivity analysis (SA) and probabilistic sensitivity analysis (PSA) to characterize uncertainty in model-based cost-effectiveness results. In SA, we vary key parameters over plausible ranges and evaluate the impact on the ICER using tornado diagrams (Supplementary Appendix 8). The set of parameters examined in SA is consistent with those included in PSA (e.g., discount rate, vaccine price and administration fee, antibody test cost, treatment costs, external force of infection, time horizon, working-age upper bound, birth-rate multiplier, and disease duration). In addition, SA considers disability weights for mild, moderate, and severe illness. In both SA and PSA, the upper bound parameter for working-age is incorporated only under the societal perspective and is not applied to the healthcare system perspective.

In the PSA, uncertain parameters are jointly sampled from prespecified distributions to perform Monte Carlo simulations for each scenario. For each iteration, costs and DALYs are recomputed to construct cost-effectiveness planes and cost-effectiveness acceptability curves (CEACs). Treatment-cost distributions are specified separately for the healthcare system and societal perspectives. Comprehensive details regarding distributional assumptions are provided in Supplementary Appendix 6.

We follow a prespecified analytic framework for the economic evaluation (Methods and Supplementary Appendix 5, 6, and 8); no separate health economic analysis plan is registered. We do not conduct a distributional (equity-focused) economic evaluation, and results are reported at the population level under healthcare system and societal perspectives. This economic evaluation is reported in accordance with the Consolidated Health Economic Evaluation Reporting Standards (CHEERS) 2022 statement.

**Results**

**Scenario-specific optimal vaccination trajectories**

Figure 2 illustrates the optimal monthly adult vaccination allocation trajectories across the three scenarios. These trajectories illustrate how vaccination effort would be preferentially allocated across age groups under each strategy to maximize health and socioeconomic benefits. In $S1$, the optimal control strategy distributes doses equally to adults in their 20s ($G_2$) and 30s ($G_3$) for approximately the first six years, followed by a seven-year period in which vaccination is concentrated in the 30s age group (Figure 2a). In $S2$, doses are evenly distributed to adults in their 40s ($G_4$) and 50s ($G_5$) for about 2.5 years, followed by approximately 7.5 years of vaccination concentrated in the 50s age group (Figure 2b). Across these two scenarios, the optimal trajectories consistently suggest initiating vaccination as early as possible and allocating a larger share of doses to relatively older target age groups. This pattern reflects the combined effects of cohort-shifted susceptibility and sharply increasing age-specific fatality, whereby vaccinating individuals closer to higher-risk ages yields larger reductions in deaths per administered dose.

In $S3$, the optimal control strategy distributes doses almost evenly across adults in their 20s–50s from 2025 through the early 2040s, effectively splitting the monthly capacity across four target age groups (Figure 2c). From the early 2040s, vaccination in the 20s ($G_2$) and 40s ($G_4$) is phased out, with allocations shifting exclusively toward the 30s ($G_3$) and 50s ($G_5$). In the final phase, the remaining vaccination effort is briefly concentrated in the 30s ($G_3$), before adult vaccination drops to zero. Overall, this trajectory reflects a strategic trade-off between minimizing deaths (by prioritizing older adults with higher mortality burden) and reducing additional program costs associated with pre-vaccination antibody testing required for adults aged ≥40 years. The resulting allocation balances epidemiological efficiency with economic feasibility under constrained adult vaccination capacity. The symptomatic infections and deaths under each optimal trajectory by scenario are provided in Supplementary Appendix 7.

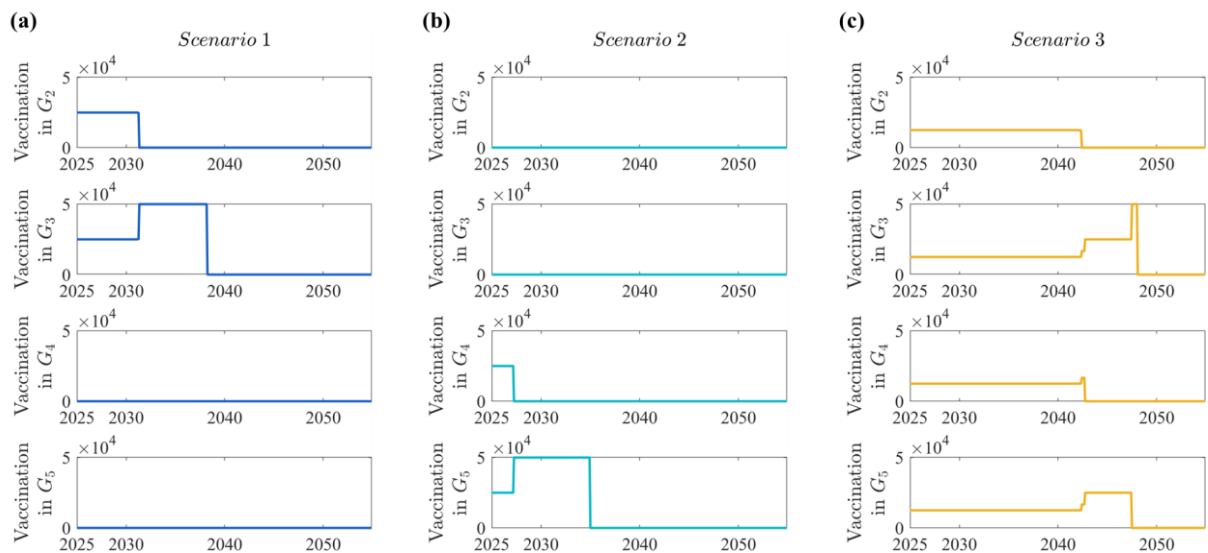

**Fig. 2** Optimal monthly vaccination allocation trajectories by age group under (a) $S1$, (b) $S2$, and (c) $S3$. Trajectories indicate how limited vaccination capacity is optimally distributed across age groups over time to minimize long-term health and socioeconomic costs under each scenario. The vertical axis shows the number of first-dose vaccinations administered per month.

**Base-case cost-effectiveness results**

The base-case cost, DALYs, and ICERs evaluated from both healthcare system and societal perspectives are presented in Table 3. The cost-effectiveness results across the three scenarios remain consistent across perspectives, with $S2$ consistently emerging as the most efficient strategy. In particular, this scenario maintains the lowest ICER under both perspectives (Table 3).

From the healthcare system perspective, $S2$ yields the lowest ICER (USD 52,166 per DALY averted), achieving a sizable reduction in DALYs (7,458 DALYs averted) with a moderate increase in costs. $S3$ produces the largest reduction in disease burden (8,605 DALYs averted) but at substantially higher incremental costs, resulting in a higher ICER than $S2$ (USD 74,863 per DALY averted). $S1$ yields a relatively small reduction in DALYs (3,557 DALYs averted) and therefore the highest ICER (USD 110,894 per DALY averted). The same ranking order is observed from the societal perspective, with $S2$ remaining the lowest ICER strategy. This consistency across analytical perspectives reinforces the robustness of $S2$ as the optimal intervention.

Table 3 Base-case cost, DALYs, and ICERs by perspectives (2025 USD, discounted at 4.5% per year). Results are shown for both healthcare system and societal perspectives. Costs and DALYs represent discounted lifetime totals (83-year evaluation horizon).

| | | Cost | Incremental cost | DALY | DALY averted | ICER |
|---|---|---|---|---|---|---|
| | | USD | | DALY | | USD/DALY |
| Healthcare system perspective | Scenario 1 ($S1$) | 537,124,476 | 394,498,172 | 14,231 | 3,557 | 110,894 |
| | Scenario 2 ($S2$) | 531,687,874 | 389,061,570 | 10,330 | 7,458 | 52,166 |
| | Scenario 3 ($S3$) | 786,837,921 | 644,211,617 | 9,183 | 8,605 | 74,863 |
| | Baseline (no additional adult vaccination) | 142,626,304 | - | 17,788 | - | - |
| Societal perspective | Scenario 1 ($S1$) | 715,791,723 | 452,406,546 | 14,231 | 3,557 | 127,172 |
| | Scenario 2 ($S2$) | 811,926,982 | 548,541,805 | 10,330 | 7,458 | 73,550 |
| | Scenario 3 ($S3$) | 1,095,056,567 | 831,671,390 | 9,183 | 8,605 | 96,647 |
| | Baseline (no additional adult vaccination) | 263,385,177 | - | 17,788 | - | - |

**Probabilistic sensitivity analysis: cost-effectiveness planes**

To evaluate the impact of parameter-level uncertainties on the simulation results, a probabilistic sensitivity analysis (PSA) is employed. The uncertain parameters and their assumed distributions are summarized in Supplementary Table 5, including vaccine acquisition and administration costs, antibody testing costs, treatment costs, vaccine-related adverse event parameters, and key economic assumptions such as the upper age limit for productivity losses. Utilizing the PSA outputs, we present cost-effectiveness planes for both perspectives based on 10,000 Monte Carlo simulations per scenario (Figure 3). Under both perspectives, $S1$ is concentrated in the region with relatively low incremental costs but comparatively small reductions in DALYs, whereas $S3$ achieves the largest DALY reductions but at substantially higher incremental costs. $S2$ clusters at moderate incremental costs while providing substantial reductions in DALYs. In the societal perspective, incremental costs for $S2$ and $S3$ are generally higher than those in the healthcare system perspective, reflecting the inclusion of additional cost components such as productivity losses and non-medical expenses. Overall, the relative positioning of the PSA point clouds is consistent with the base-case ranking, with $S2$ showing the most favorable balance between incremental costs and DALYs averted.

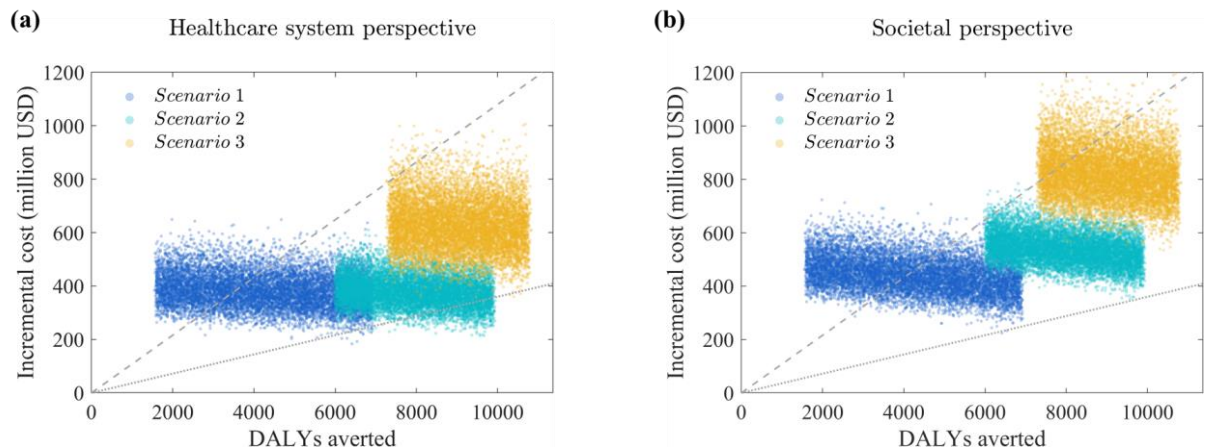

**Fig. 3** Cost-effectiveness planes for the three scenarios under (a) healthcare system perspective and (b) societal perspective. Dashed and dotted lines indicate reference ICER values corresponding to 3× and 1×GDP per capita (USD per DALY averted) for illustrative purposes only; no official WTP threshold is applied in the analysis

**Cost-effectiveness acceptability curves across willingness-to-pay values**

We further derive cost-effectiveness acceptability curves (CEACs) for both perspectives from the PSA results (Figure 4). The CEAC shows, at each willingness-to-pay (WTP) value, the probability that a given strategy is the most cost-effective option. In both perspectives, $S2$ shows the highest probability of being cost-effective across almost the entire WTP range. From the societal perspective, $S1$ briefly exceeds $S3$ within the mid-range WTP interval (WTP = 47,000–84,000 USD/DALY), indicating that $S3$, which is defined as a high cost and high effect strategy, tends to become competitive primarily when sufficiently high WTP values are assumed. Nevertheless, because $S2$ maintains the highest probability over most of the WTP range, it emerges as the most consistently preferred strategy under uncertainty. Given the absence of an official WTP threshold, we present results over a broad WTP range of 0–100,000 USD per DALY averted to reflect plausible decision-maker preferences and to assess the robustness of scenario ranking.

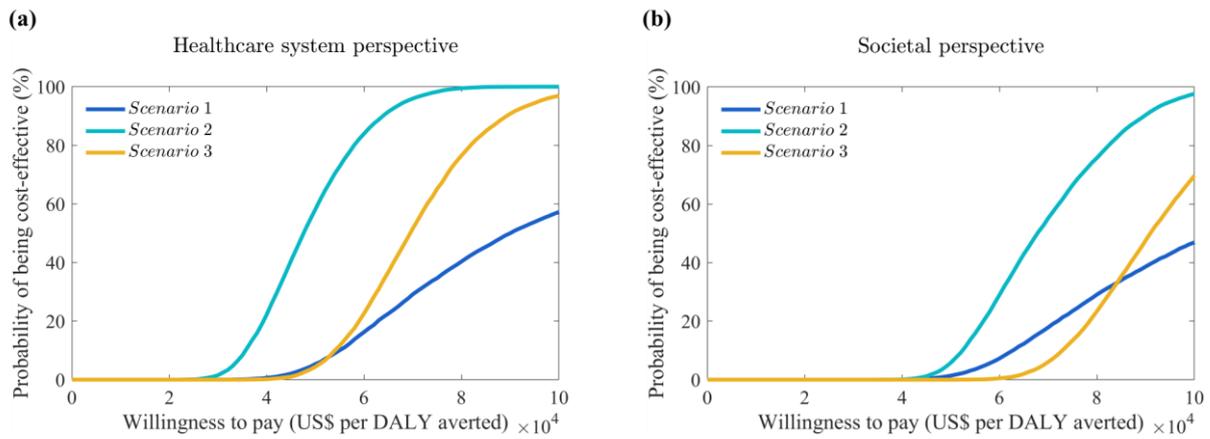

**Fig. 4** Cost-effectiveness acceptability curves (CEAC) for each scenario under (a) healthcare system perspective and (b) societal perspective. Curves show the probability that each strategy is the most cost-effective option across a range of willingness-to-pay (WTP) values. Results emphasize relative ranking and robustness across WTP values rather than reliance on a single threshold.

Taken together, these findings suggest that $S1$ is unlikely to be effectively preferred under either perspective across most WTP values considered, whereas $S3$ would be favored mainly under higher WTP values due to its high incremental costs.

**Discussion**

This study demonstrates that, under Korea's ongoing cohort shift in hepatitis A susceptibility, prioritizing adult vaccination among individuals aged 40–59 years offers the most robust and efficient strategy for reducing long-term disease burden. Across both healthcare system and societal perspectives, vaccination targeting this age group consistently achieved the most favorable balance between health gains and incremental costs, yielding the lowest ICER and the highest probability of being cost-effective under uncertainty. While broader vaccination of adults aged 20–59 years produced the largest absolute reductions in DALYs, these gains required substantially higher investment and were favored primarily at higher willingness-to-pay levels. In contrast, vaccination restricted to younger adults yielded comparatively limited health benefits. Taken together, these findings indicate that, under realistic capacity and budget constraints, prioritizing middle-aged adults provides the clearest value for money in Korea's evolving epidemiological context.

We developed an age-structured hepatitis A transmission model integrated with a DALY-based economic evaluation framework to assess the long-term impact of three additional adult vaccination scenarios. Korea does not apply a single explicit ICER threshold in health technology assessment, and international guidance has increasingly cautioned against using GDP-based rules as a universal decision criterion [24–28]. Accordingly, we presented results under both healthcare system and societal perspectives and characterized uncertainty using probabilistic sensitivity analysis (PSA) over a wide range of willingness-to-pay (WTP) values. In the base-case analysis, $S2$ (vaccination targeting adults aged 40–59 years with pre-vaccination antibody testing) produced the lowest ICER under both perspectives. This ranking was consistent under uncertainty; on the cost-effectiveness plane, $S2$ clustered in a region with substantial DALYs averted at relatively moderate incremental costs. Furthermore, CEACs showed that $S2$ had the highest probability of being cost-effective across almost the entire WTP range considered. In the absence of a formally defined national WTP threshold, the primary policy relevance of these findings lies in the stability of strategy ranking rather than in absolute ICER values. Accordingly, probabilistic dominance and relative ranking across strategies, summarized by cost-effectiveness acceptability curves, provide the main basis for interpretation.

The relative advantage of $S2$ can be interpreted in terms of the model dynamics. As the low-immunity cohort ages, susceptibility increasingly shifts toward older ages where severity and mortality risk are higher. Under a fixed monthly uptake constraint, allocating doses to adults closer to higher-fatality ages can avert substantially more deaths per dose than scenario focusing primarily on younger adults. At the same time, targeting 40–59-year-olds still covers economically active ages, supporting societal benefits through avoided productivity losses. These features help explain why $S2$ achieves a favorable balance between health benefits (DALYs averted) and incremental program costs, with benefits spanning older-age mortality and working-age illness burden.

By contrast, $S3$, the broader 20–59-year strategy, generated the largest reductions in disease burden but at substantially higher incremental costs, implying that it becomes attractive primarily when decision-makers tolerate higher WTP values or have greater budget flexibility. $S1$, targeting 20–39 years, incurred relatively low incremental costs but yielded fewer DALYs averted, resulting in the highest ICER and a consistently low probability of being the most cost-effective option under uncertainty. These results illustrate diminishing marginal returns when expanding vaccination to lower-risk age groups under constrained resources.

In terms of the broader evidence base, economic evaluations of hepatitis A vaccination have largely focused on childhood or adolescent immunization or specific risk-based programs, and fewer studies have examined long-term, adult-targeted strategies in settings where cohort shifts create aging susceptibility [29–35]. Our study contributes by integrating age-structured transmission dynamics with a long-horizon DALY-based economic evaluation, explicitly reflecting Korea's evolving immunity profile and quantifying robustness under uncertainty rather than relying on a single threshold-based conclusion.

Economic evaluation results depend on assumptions regarding the discount rate and time horizon, particularly for preventive interventions with long-term benefits such as vaccination. In this study, sensitivity analyses demonstrate that although absolute ICER values vary with these assumptions, the relative ranking of adult vaccination strategies remains unchanged across plausible ranges. This stability supports the robustness of the prioritization conclusions under Korea-specific pharmacoeconomic assumptions.

Several limitations should be noted. First, long-term projections depend on assumptions about future transmission intensity captured through the force of infection. Because the force of infection is not directly observable, forward simulations assumed that transmission intensity would remain stable at the calibrated level or vary modestly around the estimated values; however, substantial changes in exposure (e.g., hygiene, foodborne risk, travel patterns, or contact behavior) could alter the magnitude and timing of infections and deaths. Second, vaccination implementation in reality may differ from model assumptions. We represented routine (non-outbreak) feasibility via a fixed monthly uptake constraint, which implicitly requires sufficient uptake and adherence; lower acceptance could diminish coverage and reduce projected benefits. Third, we did not explicitly stratify high-risk subgroups with markedly elevated fatality (e.g., chronic liver disease). Incorporating risk stratification could increase the value of targeted approaches and may change the relative prioritization between age-based and risk-based strategies.

In summary, our findings highlight that Korea's cohort-driven aging of hepatitis A susceptibility fundamentally reshapes the cost-effectiveness of adult vaccination strategies. Aligning vaccination priorities with age groups facing rapidly increasing severity yields the greatest and most robust health gains under realistic constraints. These results underscore the importance of dynamic, context-specific economic evaluations for adult immunization policy in aging populations.

## Declarations

**Ethics approval**
Not applicable. This study used aggregated surveillance data and secondary data sources and did not involve individual-level information.

**Patient and public involvement**
Not applicable. Patients and the public were not involved in the design, conduct, or reporting of this study.

**Competing interests**
The authors declare that they have no competing interests.

**Funding**
EJ was supported by the Korea National Research Foundation (NRF) grant funded by the Korean government (MEST) (NRF-2021R1A2C100448711). EJ was also supported by 'The Government-wide R&D to Advance Infectious Disease Prevention and Control', Republic of Korea (grant number: HG23C1629). GC was partially supported by NSF grant DBI 2412115 as part of the US NSF Center for Analysis and Prediction of Pandemic Expansion (APPEX).

**Data availability**
The data sources used in this study are publicly available through the Korea Disease Control and Prevention Agency (KDCA).

**Code availability**
Model and economic evaluation code is available from the corresponding author upon reasonable request.

**Author contributions**

**Conceptualization:** Yuna Lim, Gerardo Chowell, Eunok Jung

**Formal analysis:** Yuna Lim, Gerardo Chowell, Eunok Jung

**Funding acquisition:** Eunok Jung, Gerardo Chowell

**Investigation:** Yuna Lim, Gerardo Chowell, Eunok Jung

**Methodology:** Yuna Lim

**Project administration:** Eunok Jung

**Software:** Yuna Lim

**Supervision:** Eunok Jung

**Validation:** Yuna Lim, Gerardo Chowell

**Visualization:** Yuna Lim

**Writing - original draft:** Yuna Lim

**Writing - review & editing:** Yuna Lim, Gerardo Chowell, Eunok Jung